# Inelastic Phonon Scattering in Graphene FETs

Jyotsna Chauhan and Jing Guo
Department of Electrical and Computer Engineering
University of Florida, Gainesville, FL, 32611

## ABSTRACT

Inelastic phonon scattering in graphene field-effect transistors (FETs) is studied by numerically solving the Boltzmann transport equation in three dimensional real and phase spaces ($x, k_x, k_y$). A kink behavior due to ambipolar transport agreeing with experiments is observed. While low field behavior has previously been mostly attributed to elastic impurity scattering in earlier studies, it is found in the study that even low field mobility is affected by inelastic phonon scattering in recent graphene FET experiments reporting high mobilities. As the FET is biased in the saturation regime, the average carrier injection velocity at the source end of the device is found to remain almost constant with regard to the applied gate voltage over a wide voltage range, which results in significantly improved transistor linearity compared to what a simpler model would predict. Physical mechanisms for good linearity are explained, showing the potential of graphene FETs for analogue electronics applications.

## I. Introduction

Graphene[1-8] has been one of the most rigorously studied research materials since its inception in 2004. There has been a lot of study focused on low electric field transport properties of graphene[8-12]. Many issues related to high field transport properties in graphene field-effect transistors(FETs), however, still remain unclear. Experimental[13-16] and theoretical[17-21] studies have shown that even though being a gapless semi metallic material, a graphene FET shows saturating *I-V* behaviors attributed to inelastic surface polar phonon scattering induced by gate oxide.

In this work, inelastic phonon scattering in graphene transistors is studied by numerically solving the Boltzmann transport equation (BTE) in the real and phase spaces. Modeling of inelastic surface polar phonon scattering reveals that low bias mobility is also controlled by inelastic surface polar phonon scattering in addition to elastic scattering. Good linearity is observed in high quality FETs in agreement with recent experiments and it is explained by average carrier injection velocity at the source end remaining nearly constant with the increase in applied gate bias voltage, which makes them desirable for analogue applications. Comparisons to previously reported simpler models clarify validity of those models.

To describe semiclassical transport behaviors of graphene FETs at a channel length that quantum mechanical Klein tunneling is not important, a semiclassical Boltzmann transport equation is solved self consistently with Poisson equation. Though Monte Carlo method and numerical solutions of solving BTE have been implemented in previous studies[17-21], they are limited to two-dimensional *k* space, which assumes a homogeneous material and therefore it has limitations to describe transport properties and interplay of self-consistent electrostatics and transport in a graphene transistor accurately as discussed in detail later.

## II. Simulation Approach

Top gated graphene FETs as shown in Fig.1 were simulated. The nominal device has a top gate insulator thickness of $t_{ins}=10$ nm and dielectric constant of $\varepsilon=3$, which results in a gate insulator capacitance of $C_{ins}=270$nF/cm$^2$ close to the value in a recent experiment[14]. The dielectric constant used here is close to that of hexagonal Boron Nitride (*BN*), which has been explored as one of the promising gate insulators[22] for graphene FETs .

To capture ambipolar transport properties in graphene, we considered the transport in conduction band as well as valence band. A simple linear *E-k* relation is used, which is valid in the energy range of interest,

$$E = \hbar \upsilon_F \sqrt{k_x^2 + k_y^2}, \tag{1}$$

where $\upsilon_F \approx 9.3 \times 10^7$ cm/s is the Fermi velocity in graphene[23].

The Boltzmann transport equation (BTE) is solved for a two-dimensional graphene device at the steady state. The Boltzmann Transport equations is given as[24] :

$$\frac{\partial f}{\partial t} + \frac{\vec{F}_{ext}}{\hbar} \nabla_k f + \vec{v} \nabla_r f = \frac{\partial f}{\partial t}|_{collision}, \tag{2}$$

where $f(\vec{k})$ is the distribution function, $\frac{\partial f}{\partial t}|_{collision}$ is the collision term , $\vec{F}_{ext}$ is the force on carrier due to electric field, $\vec{v}$ is the group velocity of a particular sub band and $\hbar$ is the reduced Planck's constant. Thus BTE involves a collective space containing three dimensions in real space and three dimensions in *k* or momentum space. However , since graphene is two dimensional and wide enough giving translational symmetry in the *y* direction, the problem can be easily reduced to three dimensional space of *x*, $k_x$ , $k_y$ only. Hence only three coordinates and time are needed to specify the system and BTE for graphene in three dimensional space takes the following form:

$$\frac{\partial f}{\partial t} = -\frac{\vec{F}_{ext}}{\hbar}\nabla_k f - \vec{v}_x \nabla_x f + \frac{\partial f}{\partial t}\bigg|_{collision}, \quad (3)$$

Time evolution of electronic states is described by terms $-\frac{\vec{F}_{ext}}{\hbar}\nabla_k f - \vec{v}_x \nabla_x f$. This operator is discretized on finite difference grid resulting in matrix operator. Backward or forward difference method is used depending on the direction of flux of carriers. For a charge carrier moving from left to right with positive velocity $\vec{v}_x \nabla_x f$ operator is approximated by backward difference scheme and vice versa for charge carriers moving from right to left with negative velocity. Thus $\vec{v}_x \nabla_x f$ in finite difference scheme for the carrier at $(x_i, k_{x_j}, k_{y_k})$ node is given as:

$$v_j \frac{\partial f}{\partial x} = (f_{i,j,k} - f_{i-1,j,k})\frac{v_j}{\Delta x} \text{ if } v_j > 0$$
$$= (f_{i+1,j,k} - f_{i,j,k})\frac{v_j}{\Delta x} \text{ if } v_j < 0, \quad (4)$$

where $v_j = \frac{k_{x_j}}{\sqrt{k_{x_j}^2 + k_{y_k}^2}}$ is the velocity of carriers in the transport $x$ direction at $(k_{x_j}, k_{y_k})$ node in the discretization grid and $f_{i,j,k}$ is the distribution function at $(x_i, k_{x_j}, k_{y_k})$ node.

The $\frac{\vec{F}_{ext}}{\hbar}\nabla_k f$ term includes discretization in $k_x$ space because electric field $\vec{\varepsilon}$ has only the $x$ component. So differentiation with respect to $k_y$ is eliminated. $\vec{F}_{ext} = \pm q\vec{\varepsilon}$ is the force on carrier in presence of electric field. For $\vec{\varepsilon} > 0$, electrons experience force from right to left and forward difference scheme is used and vice versa for $\vec{\varepsilon} < 0$, electrons experience force from left to right and differential operator is approximated using forward difference scheme.

Thus for electrons, the equation is solved as :

$$\frac{\vec{F_{ext}}}{\hbar}\nabla_{k_{x_j}}f = -\frac{q\mathcal{E}}{\hbar}\frac{\partial f}{\partial k_x} = (f_{i,j-1,k} - f_{i,j,k})\frac{q\mathcal{E}_i}{\hbar\Delta k_{x_j}} \; if \; \mathcal{E}_i < 0$$

$$= (f_{i,j,k} - f_{i,j+1,k})\frac{q\mathcal{E}_i}{\hbar\Delta k_{x_j}} \; if \; \mathcal{E}_i > 0, \qquad (5)$$

where $f_{i,j,k}$ is the distribution function at $(x_i, k_{x_j}, k_{y_k})$ node and $\mathcal{E}_i$ is the electric field at $x_i$ position along the channel direction.

Alternate methods of discretization of differential operators can also be used to simplify differential operators involved. After the discretization at all nodes of phase space, the differential operators in Eq. (3) takes the form:

$$-\frac{\vec{F_{ext}}}{\hbar}\nabla_{kx}f - \vec{v}_x\nabla_xf = [U]\{f^n\} + \{f_B\}, \qquad (6)$$

where $[U]$ is matrix equivalent for differential operator in $x$ and $k_x$. The $\{f_B\}$ incorporates the boundary conditions due to $\nabla_x f$ and is physically equivalent to influx of carriers at right and left contacts. At the left contact *i.e.* source contact, only positive moving carriers are responsible for influx into the device, therefore, the boundary condition at *x=0* holds true for $v_x > 0$ only.

Thereby, left contact couples to all the $k_x$, $k_y$ nodes at *x=0* and carriers are distributed according to source fermi potential.

$$f_b(x=0,k) = \frac{1}{1+(\exp(E(k)+E_c(1)-\mu_L)/k_BT))} \; for \; v_x > 0, \qquad (7)$$

Similarly, right contact couples to all $k_x$, $k_y$ nodes at *x=L* for $v_x < 0$ and carriers in these states are distributed according to drain fermi level.

$$f_b(x=L_{ch},k) = \frac{1}{1+(\exp(E(k)+E_c(N_x)-\mu_R)/k_BT))} \; for \; v_x < 0, \qquad (8)$$

where $\mu_L$ and $\mu_R$ are the source and drain fermi levels, Ec(1) and Ec($N_x$) are the conduction band edges at *x=0* and *x=L* respectively and *E(k)* is energy at every $k_x$, $k_y$ node corresponding to *x* being considered in both conduction and valence band.

For $\nabla_k f$ operator, a periodic boundary condition is used. Physically, periodic boundary conditions ensures conservation of particles and doesn't change $\{f_B\}$.

$\frac{\partial f}{\partial t}|_{collision}$ term representing collision integral is discretized and results in scattering vector $\{\hat{C}f\}$.

$$\frac{\partial f}{\partial t}_{collision} = -\sum_{\vec{k}'} S(\vec{k},\vec{k}')f(\vec{k})(1-f(\vec{k}')) + \sum_{\vec{k}'} S(\vec{k}',\vec{k})f(\vec{k}')(1-f(\vec{k})), \tag{9}$$

where $f(\vec{k})$ is the distribution function, $S(\vec{k},\vec{k}')$ is the scattering rate from $\vec{k}$ state to new state $\vec{k}'$. Thus term $-\sum_{\vec{k}'} S(\vec{k},\vec{k}')f(\vec{k})(1-f(\vec{k}'))$ represents the out scattering rate and term $\sum_{\vec{k}'} S(\vec{k}',\vec{k})f(\vec{k}')(1-f(\vec{k}))$ represents the in scattering rate. As already reported in earlier studies, low field mobility is controlled by elastic phonon scattering while saturation behavior is largely dominated by inelastic surface polar phonon scattering. So as part of study, elastic phonon scattering is modeled with $\lambda$ as the fitting parameter, where $\lambda$ is the mean free path (mfp) of carrier in presence of elastic scattering. The scattering rate for elastic scattering is calculated as:

$$S(\vec{k},\vec{k}') = \frac{\upsilon_F \times E(\vec{k})}{\lambda}, \tag{10}$$

where $E(\vec{k})$ is the energy of initial state as there is no exchange of energy due to elastic scattering so energy remains the same and $\lambda$ is used as fitting parameter.

The surface polar phonon scattering rate is calculated using Fermi's golden rule[17,25,26]. For inelastic surface polar phonon emission process, the electron in conduction band can jump to valence band or stay in conduction band while the electron in valence band stays only in valence band within modeled energy range as shown in Fig.2(a). In case the electron in valence band jumps outside the considered energy range by emitting a phonon the transition is prohibited.

For surface polar phonon absorption process, the electron in conduction band stays within conduction band. In case , transition results in final energy outside the modeled range , it is prohibited. However, the electron in valence band on absorption can stay in valence band or jump to conduction band as shown in Fig.2(b).

Isotropic scattering is assumed for both elastic as well as surface polar phonon scattering. Hence, *E-k* space is a constant energy sphere and all states lying on constant energy sphere are equally probable. This leads to the assumption that initial state $\vec{k}$ is coupled to all the states lying on constant energy sphere $E(\vec{k}')$ with an equal scattering rate. The total scattering rate from $\vec{k}$ to $\vec{k}'$ is thus, distributed equally among all the states satisfying $E(\vec{k}')$.

After including scattering and all discretization terms the final BTE can be written as:

$$\frac{\{f^{n+1}\}-\{f^n\}}{dt} = [U]\{f^n\} + \{f_B\} + \{\hat{C}f\}, \qquad (11)$$

Where $\{f^{n+1}\}$ and $\{f^n\}$ are the distribution functions at *n+1* and *n* time steps. To reduce computational cost, the device is assumed to be in steady state such that $\frac{\{f^{n+1}\}-\{f^n\}}{dt}$ can be approximated as zero and equation further simplifies as:

$$[U]\{f^n\} + \{f_B\} + \{\hat{C}f\} = 0, \qquad (12)$$

To start with the simulation, initial guess of distribution function $\{f^0\}$ is calculated assuming ballistic transport conditions which gives

$$\{f^0\} = -[U^{-1}]\{f_B\}, \tag{13}$$

Thus, $\{f^0\}$ is the equilibrium ballistic distribution function. The BTE is then solved self consistently with non linear Poisson equation using Newton Raphson method.

**III.    Results and Discussion**

The study analyzes the characteristic of the graphene FETs on a micrometer channel length regime. The simulation captures the ambipolar $I_D$-$V_D$ behavior observed in graphene FETs which single band Monte Carlo simulations fails to capture. The rest of section addresses the effect of elastic scattering and inelastic surface polar phonon scattering on the transport behavior of graphene FETs. The last section explains the nearly constant transconductance observed experimentally  in graphene FETs which makes it good potential for analogue applications. Comparison to previously developed simpler models clarifies the validity of these models.

We start by simulating $I_D$-$V_D$ characteristics of graphene FETs. Figure 1 shows the schematic of the device used in our simulations. Top gated graphene FET with two dimensional graphene as the channel is simulated. A linear *E-k* for graphene is assumed in two band simulations. The graphene channel is wide enough so that there exists translational symmetry in the *y* direction, thus helping to reduce the phase space in BTE to ($k_x, k_y, x$). The device is simulated for $C_{ins}=270$ nF/cm$^2$ with Boron Nitride(BN) as the dielectric which has been predicted to be one of the promising dielectrics for graphene FETs in recent experiments[14,22]. All the simulations have been done for channel length $L_{ch}=$ *1*μm.

The current $I_D$ vs. drain voltage $V_D$ relation is studied for 1μm device. The $I_D$   - $V_D$

curve in Fig.3(a) shows the characteristic kink due to dirac point entering the drain, thereby leading to ambipolar transport behavior in device. This behavior is in agreement with experimental results[13]. Figure 3(b) and Figure 3(c) explains the onset of ambipolar transport near kink drain voltage by plotting electron ( left side plot ) and hole (right side plot) distribution function near drain end of the channel at $V_D = 0.25$V (before the kink) and $V_D = 0.7$V (after the kink) respectively. As shown in Fig.3(b) at $V_D = 0.25$V before the onset of kink, the electron distribution function is quite dominant while hole distribution function is almost close to zero. The current at this point is thus carried largely by electrons and contribution due to holes to the total current can be neglected. However as shown in Fig.3(c) at $V_D = 0.7$V after the kink, both electron and hole distribution functions become equally prominent. This explains the ambipolar behavior in the region following the kink.

The effect of scattering on mobility is studied next. It has already been observed that elastic impurity scattering and inelastic surface polar phonon scattering due to gate dielectric has been dominant scattering mechanisms in controlling the transport behavior in graphene. The low field behavior has been attributed to elastic scattering in many earlier studies [10-12,20]. Figure 4(a) shows the low field conductance in presence of elastic scattering. The conductance varies almost linearly with increase in $\lambda$, which is used as a fitting parameter in our simulations. Figure 4(b) shows the mobility for different values of $\lambda$ as a function of $V_G$. The mobility increases with the increase in value of $\lambda$. The mobility remains almost constant with increase in gate overdrive voltage because low bias conductance for elastic scattering varies linearly with gate overdrive voltage i.e. $G \propto E_f^2 \propto (V_G - V_T)$. Then we plotted the conductance and mobility values in presence of inelastic surface polar phonon scattering for different Froehlich coupling constants (material parameter) at $\hbar\omega=40meV$ with elastic scattering completely turned off . It is observed that low bias conductance as shown in Fig.5(a) and mobility as shown in Fig.5(b), calculated in presence of inelastic surface polar phonon scattering are strongly influenced by the inelastic surface polar phonon scattering in graphene. Thus even at low bias transport, considering influence of surface polar phonon scattering is quite important as the values of conductance and mobility in presence of inelastic surface polar phonon scattering are closer to the values calculated for elastic scattering. Hence, in a device when both mechanisms are present we get the mobility behavior controlled by both scattering mechanisms as per Matthiessen's rule.

Figure 6(a) shows the simulated $I_D$ and $V_D$ results. In the simulation, the elastic scattering is turned off by choosing large value of elastic scattering mean free path. It indicates the device *I-V* characteristics in the presence of only inelastic phonon scattering. We also performed simulations by turning on the elastic scattering. If the elastic scattering mfp is limited by acoustic phonon scattering, which is considerably longer than the scattering mfp by inelastic phonons modeled here, the *I-V* characteristics shows negligible difference as shown in Fig. 6(a). The $I_D$-$V_D$ characteristics for $\lambda_{elastic}=2$μm and *200*μm defined at the electron energy of $E=0.1$eV for elastic scattering remain almost same. The above condition requires a high quality graphene transistor where elastic defect scattering and screened charge impurity scattering are weak.

Figure 6(b) plots the average carrier injection velocity (dashed line) at the beginning of channel *i.e.* near the source end, as a function of the gate voltage $V_G$, which is found to remain approximately constant. It is also interesting to compare the detailed numerical simulations to previously developed simple models for saturation velocity. The simulated behavior is in contrast to saturation velocity (line with diamonds) behavior which states that[13]

$$v_{sat} = v_f \frac{\hbar\omega}{E_f}, \tag{14}$$

and thereby decreases with increase in $V_G$.

The difference is due to two reasons. First, the application of the above equation requires $E_f$ to be significantly larger than $\hbar\omega$, which is not satisfied here. Second, the average carrier velocity does not fully reach the saturation velocity value before the turn-on of ambipolar transport current, especially at low gate overdrive voltages.

We also plotted the average carrier velocity (line with crosses) for homogenous graphene material by solving BTE in just two dimensional space ($k_x$, $k_y$) using same operating conditions as calculated at the source end of the device in previous simulations. It is found that results for the graphene FET simulations agree with homogenous material simulations within 16 %. In running all the simulations, it is found that the device does not completely reach saturation regime before turn on of ambipolar transport. To illustrate this point, we also plotted the saturation velocity (line with squares) calculated by solving BTE in two dimensional space ($k_x$, $k_y$) for the homogenous material to see how much offset is the device from saturation regime. The saturation velocity plotted above is the highest velocity achieved by homogenous material[19]. It was observed that average carrier injection velocity of carriers at source end (dashed line) is below the saturation velocity (line with squares) of the homogenous material. Hence, it confirms that the device doesn't fully reach velocity saturation before the turn on of ambipolar transport.

In order to investigate the linearity of the transistor, we plotted the transconductance as a function of the gate voltage as shown in Fig.6(c). The line with asterisks, which shows the numerical BTE solution, indicates that the transconductance is nearly constant over a wide applied gate voltage range, indicating excellent linearity of graphene transistors in the saturation drain current regime. On the other hand, a simpler model without numerical solution of the BTE would have predicted much worse linearity, as shown by the line with squares curve in Fig.6(c), which plots $g_m = C_g v_{sat} = C_g v_f \frac{\hbar \omega}{E_f}$ as function of gate voltage. The simpler model (line with squares) indicates that the transconductance increases as the gate voltage decreases, mostly because the carrier saturation velocity is inversely proportional to Fermi energy level. However, almost constant injection velocity and $g_m$ is observed in the device. This is due to the fact that device doesn't reach complete saturation before turn on of ambipolar transport. Figure 6(d) shows the carrier velocity at various gate bias points for the homogenous graphene by solving BTE in two dimensional k space ($k_x$ and $k_y$). On the same plot, the velocity at the source end of the device at $V_G = 1.0V, 1.5V, 2.0V$ and $2.5V$ is highlighted with bold marker points. As can be seen the device is well below the saturation regime and the carrier injection velocity at the source end of the device over the gate bias from 1.0 to 2.5 V remains almost constant. Since $g_m = C_g v_{injection}$, it also remains constant with $V_G$ as seen in Fig. 6(c).

A numerical solution of the BTE as described in this work, therefore, is necessary for obtaining much accurate prediction of transistor *I-V* characteristics and clearly explaining excellent linearity observed in recent high-quality experimental devices with BN insulator.

High Frequency behavior of graphene FETs is also studied using quasi static treatment[27,28]. The intrinsic gate capacitance, $C_g$ and the transconductance, $g_m$, are computed by calculating derivatives of the charge in the channel and the drain current numerically at slightly different gate voltages,

$$C_g = \left.\frac{\partial Q_{ch}}{\partial V_G}\right|_{V_D}, \quad g_m = \left.\frac{\partial I_D}{\partial V_G}\right|_{V_D}, \quad (15)$$

The intrinsic cut-off frequency is computed as,

$$f_T = \frac{1}{2\pi}\frac{g_m}{C_g} \quad (16)$$

The intrinsic cut off frequency $f_T$ (line with asterisks) is computed by running self consistent simulations at $V_D=0.5V$ and $V_G=1.1\ V$. The simulated $f_T$ increases with increase in phonon energies as manifested in Fig.7. This is due to increase in saturation current with the increase in surface polar phonon energy. However, the increase is not linear. Also,

$$f_T = \frac{1}{2\pi}\frac{<v>}{L_{ch}}, \quad (17)$$

where $<v>$ is the average velocity of carriers, $L_{ch}$ is the length of the channel.

If we input $<v> = v_{sat} = v_f \frac{\hbar\omega}{E_f}$ assuming the device is in saturation, we get

$$f_T = \frac{1}{2\pi}\frac{v_f}{L_{ch}}\frac{\hbar\omega}{E_f}, \quad (18)$$

where $v_F \approx 9.3\times 10^7$ is the Fermi velocity in graphene, $\hbar\omega$ is the surface polar phonon energy and $E_f$ is the Fermi level in graphene depending on the carrier density.

The intrinsic cut off frequency $f_T$ calculated using the above equation (with squares) is also plotted in Fig.7 and shows a linear increase with surface polar phonon energy due to inverse dependence on $E_f$. The difference is due to the fact that as $\hbar\omega$ increases, Eq. (14) breaks down and the carrier velocity does not increase as fast as a proportional relation. For $\hbar\omega=20$meV, the Eq.(14) remains valid and the value calculated is almost equal to simulated value of $f_T$. To summarize, if an applied gate voltage results in a source Fermi Level $E_f$ with reference to the Dirac point at the source end of the channel, then increasing phonon energy $\hbar\omega$ leads to increase of intrinsic cut off frequency $f_T$ if $|E_f| > \hbar\omega$ which indicates the possibility of improving the high frequency performance by phonon engineering. However, if $\hbar\omega$ becomes comparable or larger than $|E_f|$, the $f_T$ becomes insensitive to phonon energy.

**IV.  Conclusions**

In Summary, we study the characteristics of graphene FETs in presence of elastic scattering and surface polar phonon by running two band, self consistent simulations of Boltzmann transport equation. A characteristic kink behavior already seen in experiments is observed in simulations. It is found that low bias transport regime is affected by both elastic as well as inelastic phonon scatterings. The average velocity at source end remains constant with gate bias as the device doesn't reach full velocity saturation before turn on of ambipolar transport which explains for the nearly constant transconductance seen experimentally. A simple relation, $v_{sat} = v_f \frac{\hbar\omega}{E_f}$, tends to overestimate the saturation velocity as well as intrinsic cut off frequency of the graphene FETs under certain bias conditions. The intrinsic cut off frequency $f_T$ can be improved by phonon engineering if the condition $|E_f| > \hbar\omega$ is satisfied. However, numerical Boltzmann transport equation being detailed approach captures the device physics of graphene FETs in micrometer regime and explains for the characteristics observed experimentally.

## Acknowledgement

The authors would like to thank Dr. Eric Snow of Naval Research Lab for discussions on linearity of graphene FETs, and Prof. Ken Shepard and Inanc Meric of Columbia University for technical discussions. The work is supported by NSF, ONR and ARL.

**Figure Captions**

Fig.1 a) Cross section of the device structure simulated. b) Schematic representation of the E-k diagram of graphene in which two inequivalent valleys are shown.

Fig. 2 Schematic sketches of (a) inelastic phonon emission and (b) inelastic phonon absorption for electrons in conduction and valence bands of graphene.

Fig.3 a) The current $I_D$ vs $V_D$ curve for $L_{ch}=1$μm at $V_G=0.6$V. A characteristic kink is observed due to the ambipolar transport behavior in graphene FETs, in agreement with experiments[13]. Simulated electron (left) and hole (right) distribution functions in the channel near drain side at (b) $V_D=0.25$V (before kink) and (c) $V_D=0.7$V (after kink).

Fig.4 a) The simulated low field channel conductance for elastic scattering as a function of gate voltage $V_G$ at $V_D=0.1$V with $\lambda=40$nm (line with squares), $\lambda=80$nm (line with asterisks), $\lambda=120$nm (line with diamonds), which is defined at the electron energy of $E=0.1$eV for elastic scattering. (b) Mobility as function of gate voltage $V_G$ in presence of elastic scattering at $V_D=0.1$V. Inelastic surface polar phonon scattering is turned off in all these simulations.

Fig. 5 a) The Low field channel conductance for surface polar phonon scattering as a function of gate voltage $V_G$ at $V_D=0.1$V with Froehlich coupling constant of $1.0$eV (line with squares), $1.5$eV (line with asterisks), $2.0$eV (line with diamonds) at $\hbar\omega=40$meV. (b) Mobility as a function of gate voltage $V_G$ in presence of surface polar phonon scattering at $V_D=0.1$V. Elastic scattering is absent in all these simulations.

Fig.6 a) The $I_D$-$V_D$ characteristics at $V_G=0.8, 1.5, 1.8, 2.3$ V (bottom to top) with $\lambda_{elastic}=2$μm (line with squares) and $\lambda_{elastic}=200$μm (line with triangles) defined at the electron energy of $E=0.1$eV for elastic scattering, where the channel length is $L_{ch}=1$μm, SPP energy $\hbar\omega=40$meV, and gate insulator capacitance $C_{ins}=270$nF/cm$^2$. b) The simulated average carrier injection velocity as a function of the gate voltage V$_G$(dashed line) at $V_D=0.6$ V. For comparison, the following three velocities computed from simpler models are also plotted. (i)The saturation velocity (line with squares) computed from a 2D BTE solver in the ($k_x$,$k_y$) space, (ii) the average velocity computed from the 2D BTE solver in the ($k_x$,$k_y$) space at the same electric field as that at the beginning of the transistor channel at $V_D=0.6$V(line with crosses), and (iii) The saturation velocity(line with diamonds) computed by a simple model in Eq. (14),

$$v_{sat}=v_f\frac{\hbar\omega}{E_f}$$. c) The simulated transconductance (line with asterisks) $g_m$ at $V_D=0.6$V vs. the gate voltage $V_G$. For comparison, the line with squares shows $g_m$ obtained as the gate capacitance times the saturation velocity computed by Eq. (14) at $V_D=0.6$V. d) Velocity Vs Electric Field at $V_G=1.0$V(line with a triangle), $1.5$V( line with a circle), $2.0$V(line with a diamond) and $2.5$V(line with a square) at $V_D=0.6$V (before onset of ambipolar regime).

Fig. 7 The simulated intrinsic cutoff frequency $f_T$(line with asterisks) vs. phonon energy ($\hbar\omega$) curve shows an increase in intrinsic cut off frequency with the increase in phonon energy. The plot also shows the intrinsic cutoff frequency calculated using Eq.(18) under the same operating conditions which shows almost linear dependence on $\hbar\omega$. The intrinsic cut-off frequencies are computed at $V_G=1.1$V and $V_D=0.5$V.

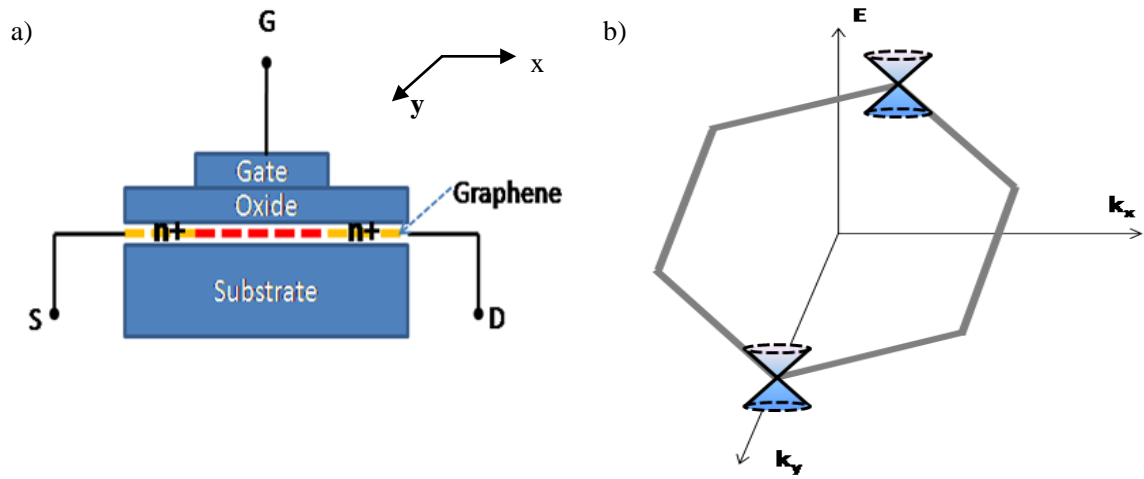

Fig.1 a) Cross section of the device structure simulated. b) Schematic representation of the E-k diagram of graphene in which two inequivalent valleys are shown.

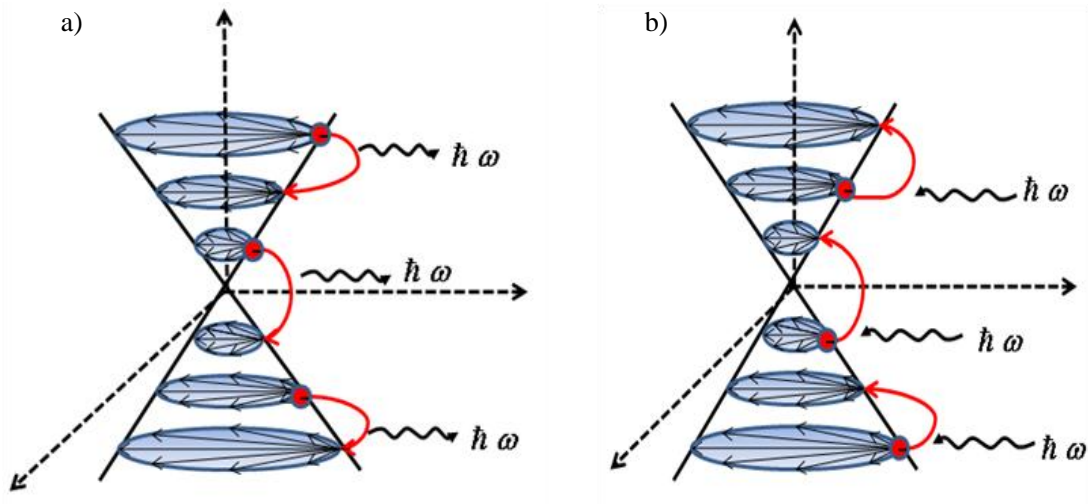

Fig.2 Schematic sketches of (a) inelastic phonon emission and (b) inelastic phonon absorption for electrons in conduction and valence bands of graphene

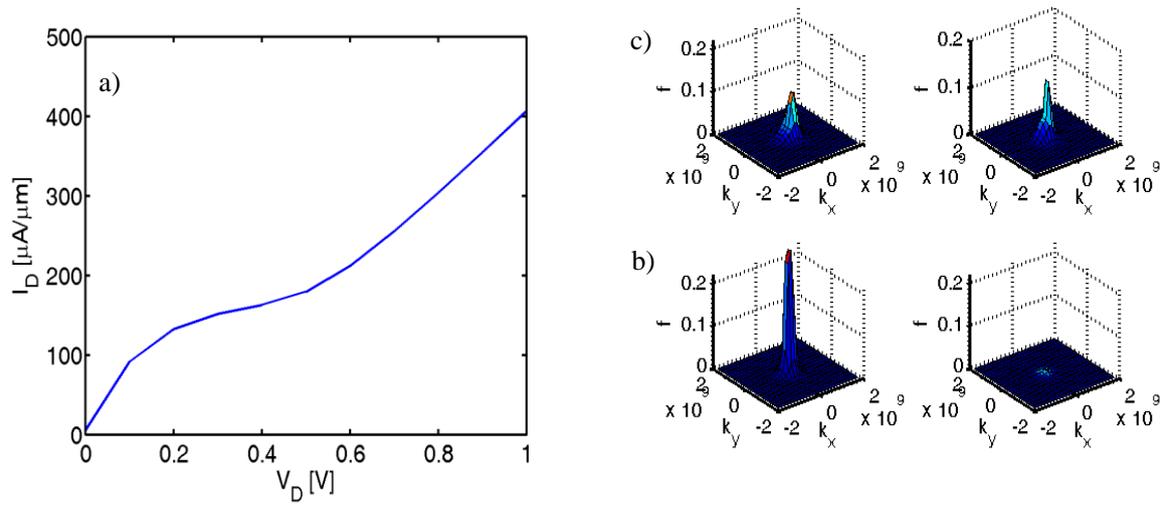

Fig. 3 a) The current $I_D$ vs $V_D$ curve for $L_{ch}=1$μm at $V_G= 0.6$V. A characteristic kink is observed due to the ambipolar transport behavior in graphene FETs, in agreement with experiments[13]. Simulated electron (left) and hole (right) distribution functions in the channel near drain side at (b) $V_D =0.25$V (before kink) and (c) $V_D =0.7$V (after kink).

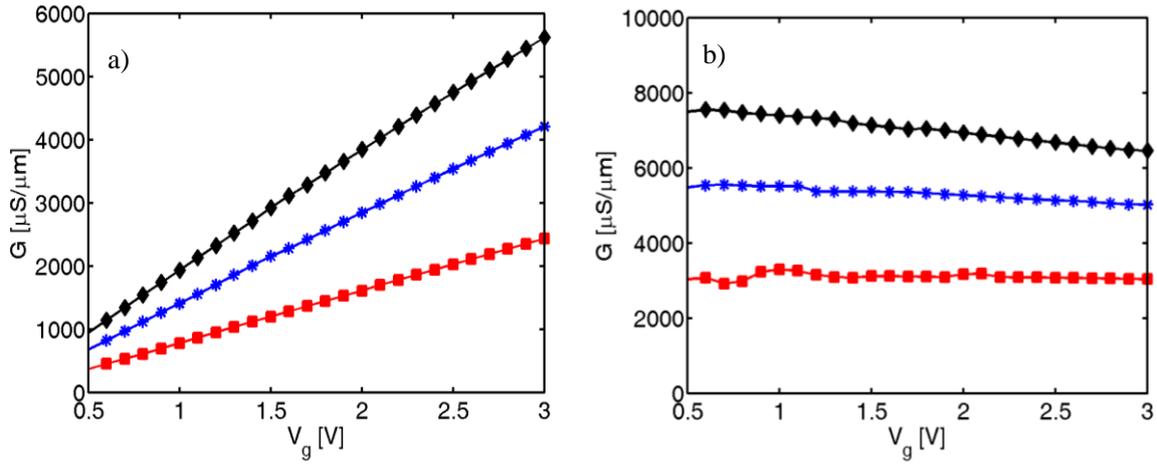

Fig.4 a) The simulated low field channel conductance for elastic scattering as a function of gate voltage $V_G$ at $V_D=0.1$V with $\lambda=40$nm(line with squares), $\lambda=80$nm(line with asterisks), $\lambda=120$nm ( line with diamonds), which is defined at the electron energy of $E=0.1$eV for elastic scattering. (b) Mobility as function of gate voltage $V_G$ in presence of elastic scattering at $V_D=0.1$V . Inelastic surface polar phonon scattering is turned off in all these simulations.

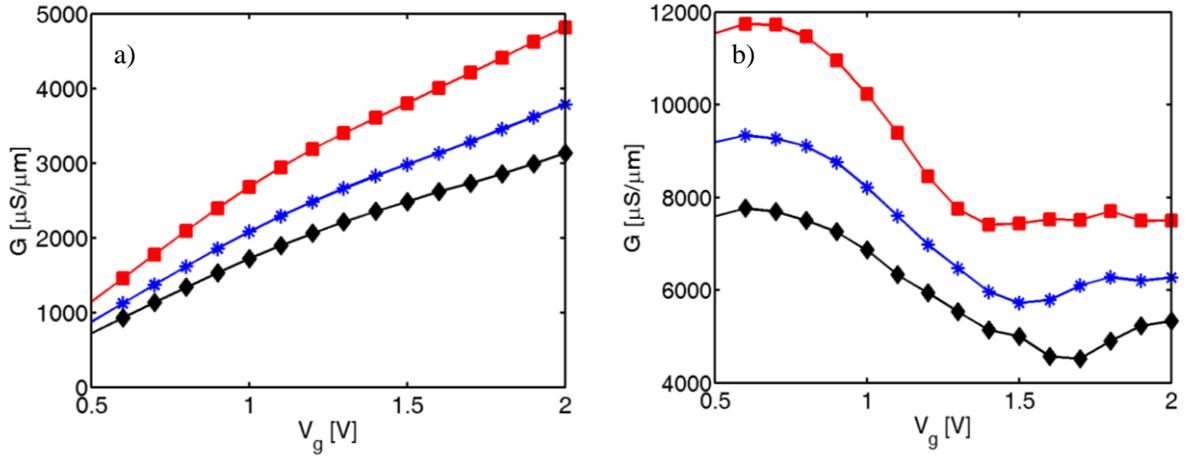

Fig. 5 a) The low field channel conductance for surface polar phonon scattering as a function of gate voltage $V_G$ at $V_D=0.1$V with Froehlich coupling constant of $1.0$eV (line with squares), $1.5$eV (line with asterisks), $2.0$eV (line with diamonds) at $\hbar\omega=40$meV. (b) Mobility as function of gate voltage $V_G$ in presence of surface polar phonon scattering at $V_D=0.1$V. Elastic scattering is absent in all these simulations.

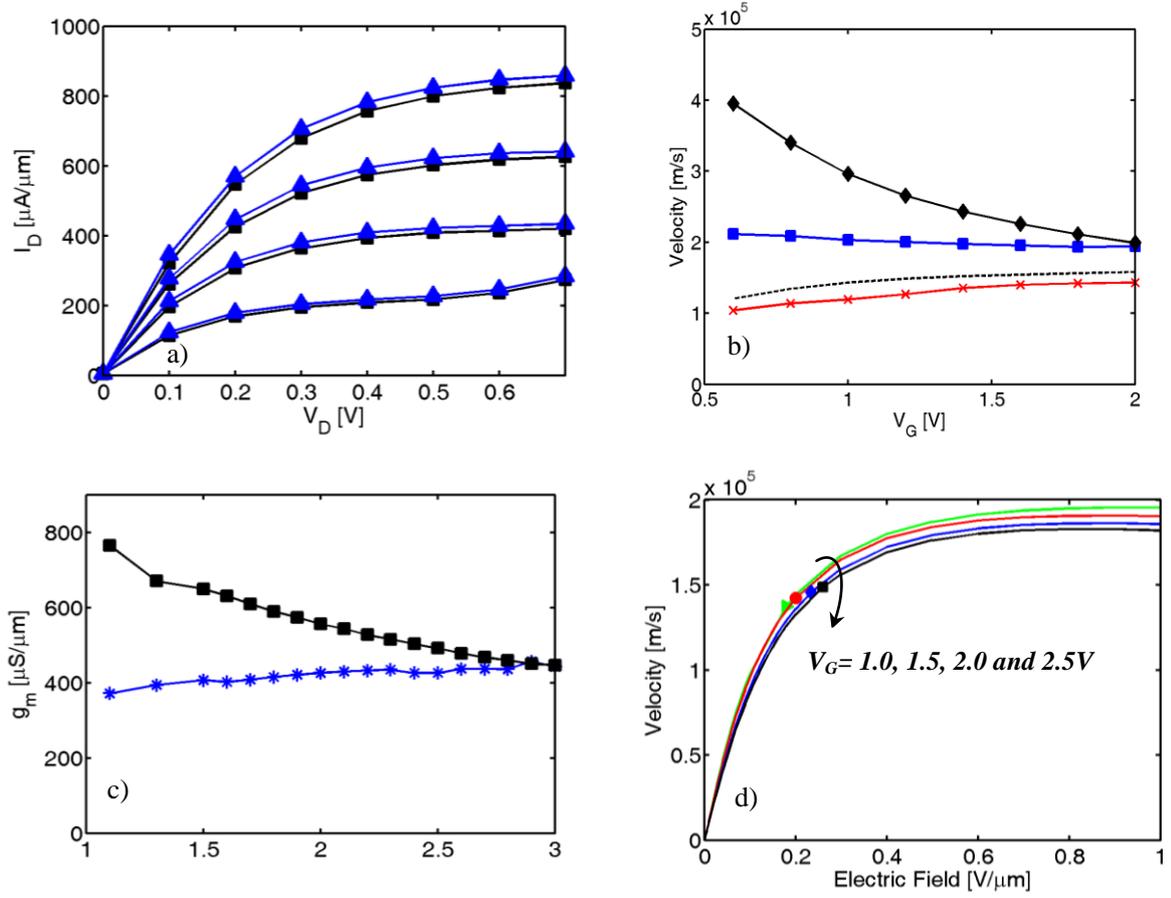

Fig.6 a) The $I_D$-$V_D$ characteristics at $V_G$=0.8,1.5,1.8,2.3 V (bottom to top) with $\lambda_{elastic}$=2μm (line with squares) and $\lambda_{elastic}$=200μm (line with triangles) defined at the electron energy of E=0.1eV for elastic scattering, where the channel length is $L_{ch}$ =1μm, SPP energy $\hbar\omega$ =40meV, and gate insulator capacitance $C_{ins}$=270nF/cm². b) The simulated average carrier injection velocity as a function of the gate voltage $V_G$(dashed line) at $V_D$=0.6 V. For comparison, the following three velocities computed from simpler models are also plotted. (i)The saturation velocity (line with squares) computed from a 2D BTE solver in the ($k_x$,$k_y$) space, (ii) the average velocity computed from the 2D BTE solver in the ($k_x$,$k_y$) space at the same electric field as that at the beginning of the transistor channel at $V_D$=0.6V(line with crosses), and (iii) The saturation velocity(line with diamonds) computed by a simple model in Eq. (14), $v_{sat} = v_f \dfrac{\hbar\omega}{E_f}$. c) The simulated transconductance (line with asterisks) $g_m$ at $V_D=0.6$V vs. the gate voltage $V_G$. For comparison, the line with squares shows $g_m$ obtained as the gate capacitance times the saturation velocity computed by Eq. (14) at $V_D=0.6$V. d)Velocity Vs Electric Field at $V_G=1.0$V(line with a triangle), *1.5*V( line with a circle), *2.0*V (line with a diamond) and *2.5*V(line with a square) at $V_D=0.6$V (before onset of ambipolar regime).

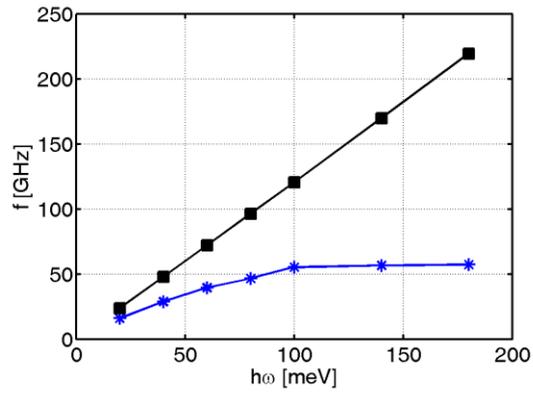

Fig.7 The simulated intrinsic cutoff frequency $f_T$ (line with asterisks) vs. phonon energy ($\hbar\omega$) curve shows an increase in intrinsic cut off frequency with the increase in phonon energy. The plot also shows the intrinsic cutoff frequency calculated using Eq.(18) under the same operating conditions which shows almost linear dependence on $\hbar\omega$. The intrinsic cut-off frequencies are computed at $V_G = 1.1$V and $V_D = 0.5$V.